\documentstyle [aps,epsfig,amsfonts,color]{revtex}

\definecolor{grey0.9}{gray}{0.9}
\definecolor{grey0.95}{gray}{0.95}
\definecolor{grey0.97}{gray}{0.97}
\definecolor{grey0.99}{gray}{0.99}

 \def\(({\left(}
 \def\)){\right)}
\def\[[{\left[}
\def\]]{\right]}

\newcommand \ee{\end{equation}}
\newcommand \be{\begin{equation}}
\def\bi{\bibitem}
\newcommand \bea {\begin{eqnarray} \nonumber }
\newcommand \eea {\end{eqnarray}}

\def \si {s_i}
\def \sih {\hat{s}_i}
\def \sid {\dot{s}_i}

\begin{document}

\title{The $p$-spin spherical spin glass model}

\author{A. Barrat \footnote{email~:barrat@ictp.trieste.it}}
\address{International Center for Theoretical Physics,
Strada Costiera 11,
34100 Trieste,
Italy}

\maketitle
\begin{abstract}
This review presents various aspects of a mean-field
spin glass model known as the p-spin spherical spin glass model,
which has raised a lot of interest in the study of spin glasses,
and also for its possible links with a mean-field theory of structural
glasses.

This preprint contains no new results and is therefore not intended
to be published, but its aim is to present a collection of results
and formulas concerning this very rich model.

It is in fact the english translation of one of the chapters of
my PhD thesis (``Quelques aspects de la dynamique hors d'equilibre
des verres de spin'', ``Some aspects of the out of equilibrium dynamics
of spin glasses''). A postscript version (in french)
of this PhD thesis will soon be available at 
http://www.lpt.ens.fr .
\end{abstract}

\vskip 1cm

The recent developments in the theory of spin glass dynamics
have made clearer the similarity of 
behaviour in
spin glasses and in glasses \cite{fraher,bocukumez}. In
this context it seems at the moment that a certain
category of spin glasses, those which are described
by a so called one step replica symmetry breaking 
(RSB) transition \cite{mezparvir}, are good candidate models for a
mean field description of the glass phase 
\cite{thirum87,wolynes,parisi_verre}.
In these systems the presence of metastable states
generates a purely dynamical transition
(which is supposed to be rounded in finite dimensional
systems \cite{thirum87,wolynes,parisi_verre})
 at a temperature $T_d$ higher than the
one obtained within a theory of static equilibrium, $T_s$. 
 
The spherical p-spin spin glass introduced in
\cite{crisom92,crihorsom} is an interesting example of this category.
It is a simple enough system in which the metastable
states can be defined and studied by the TAP method \cite{TAP}.
Besides, its relaxationnal Langevin dynamics was shown to
display the interesting behaviour known as aging \cite{cukuprl}.

The model is defined by the Hamiltonian
\begin{equation}
H =  - \sum_{1 \leq i_1 < i_2 \cdots < i_p \leq N}
J_{i_1 i_2 \cdots i_p} s_{i_1}s_{i_2}\cdots s_{i_p}
\end{equation}
with $p \ge 3$, 
where the couplings are gaussian, with zero mean and variance
$p!/(2N^{p-1})$.
The spins, instead of being restricted
to the values $+1$ or $-1$ (Ising spins),
are real variables, with the global
constraint $\sum_{i=1}^N s_i^2 = N$~: the system is less frustrated,
but this simplification allows for a more complete analytical 
treatment, and the model still displays a
very interesting behaviour.

In this small review are presented statical and dynamical aspects
of this model. Numerous formulas are displayed in appendices.

\section{Statics}

\subsection{Replica method}

This study, made by Crisanti and Sommers \cite{crisom92},
shows a transition at a temperature $T_s$, between a high
temperature replica symmetric phase, and a low temperature phase
with  one step of replica symmetry breaking (see appendix A.1)~: 
at low temperature,
the Boltzmann measure is dominated by a small number of pure states.

The static transition temperature is given by
\cite{kurparvir}~:
\be
T_S = y \sqrt{ \frac{p}{2y} } (1-y)^{\frac{p}{2} -1} ~,
\ee
where
\be
\frac{2}{p} = -2y \frac{1-y +\ln y }{(1-y)^2} .
\ee

\subsection{TAP equations}

The TAP (Thouless-Anderson-Palmer, \cite{TAP})
equations are equations on the local magnetizations 
$m_i = \langle s_i \rangle$. They were derived by Rieger \cite{riegertap} 
for the p-spin model with Ising spins~; 
they can be derived through a variational principle
on the $m_i$, from a free energy $f\((\{m_i \} \))$. In the
spherical case, this free energy was obtained by various authors
\cite{kurparvir,crisom95}~; in appendix A.2 we propose another derivation
by the cavity method \cite{mezparvir}.

The free energy $f\((\{m_i \} \))$ is best written in terms of radial
and angular variables, $q$ and  $\hat{s}_i$ (with
$m_i = \sqrt{q} \hat{s}_i$), in the form \cite{kurparvir}~: 
\be
f(\{m_i \} )=q^{\frac{p}{2}} E^0(\{ \hat{s}_i \} )
 - \frac{T}{2}\ln(1-q) -
\frac{1}{4T}[(p-1)q^p - pq^{p-1} +1] \ ~;
\label{ftap}
\end{equation}
where the angular energy is~:
\be
E^0 (\{ \hat{s}_i \} )
\equiv - \frac{1}{N}
\sum_{1\leq i_1<\cdots<i_p\leq N}\, J_{i_1,\ldots,i_p}\,
       \hat{s}_{i_1}\cdots \hat{s}_{i_p} \ .
\ee

At zero temperature the TAP states 
are just unit vectors which minimize the
angular energy $E^0$. There actually exist such states for
 $E^0 \in [E_{min}, E_c=-\sqrt{2(p-1)/p}$]\cite{crisom95}. The
states with $E^0=E_{min}$ correspond to the RSB solution.

We denote by $\hat{s}_i^\alpha$ one zero temperature state,
of energy $E^0_\alpha$.
The free energy per spin $f\((\{m_i \} \))$ depends on the
$\hat{s}_i$ only through  $E^0_{\alpha}$. We see therefore that each state
$\hat{s}_i^\alpha$, gives rise at finite
temperature $T$ to one TAP state $\alpha$ given by~:
\be
m_i^\alpha= \sqrt{q (E^0_\alpha,T )} \hat{s}_i^\alpha \ ,
\label{alph}
\ee
where $q (E^0_\alpha,T )$ is given in appendix, equation
(\ref{qtap}).

When changing the temperature, one can follow the 
metastable states which keep the same angular direction (\ref{alph}); their
order in free energy or energy, at fixed T, is the same as their
order in $E^0$~: if $E^0_\alpha > E^0_\beta$, 
$E_\alpha (T) > E_\beta (T)$.

When raising $T$, 
a state disappears at a temperature $T_{max}(E^0)$ (where the equation
defining $q (E^0_\alpha,T )$ has no more solutions)~;
$T_{max}(E^0)$ is a decreasing function of $E^0$~;
the most excited states, with $E^0=E_c$,
disappear first at $T_{max}(E_c)$,
and the lowest
at $T_{max}(E_{min}) \equiv T_{TAP}$. Above $T_{TAP}$,
the only remaining state is the paramagnetic one with $q=0$ and
free energy $F_{para}=-1/(4 T)$.

To complete the description of metastable states at any temperature, 
one only needs the density
of states $\rho(E^0)$ with an angular energy $E^0$. This
has been computed in \cite{crisom95}; the multiplicity is exponentially
large, giving a finite complexity density $s_c^0(E^0)$, defined
as~:
\be
s_c^0(E^0)= \lim_{N \to \infty} {\log \rho(E^0) \over N} \ .
\ee
At finite temperature, for a free energy  $F$ between
the free energy of the lowest (with $E^0=E_{min}$) and the
highest (with $E^0=E_c$) TAP states, we have therefore an extensive
value for $S_c(F,T)$, the logarithm of the number of TAP states 
with free energy $f$ at temperature $T$.

\begin{figure}
\centerline{\hbox{
\epsfig{figure=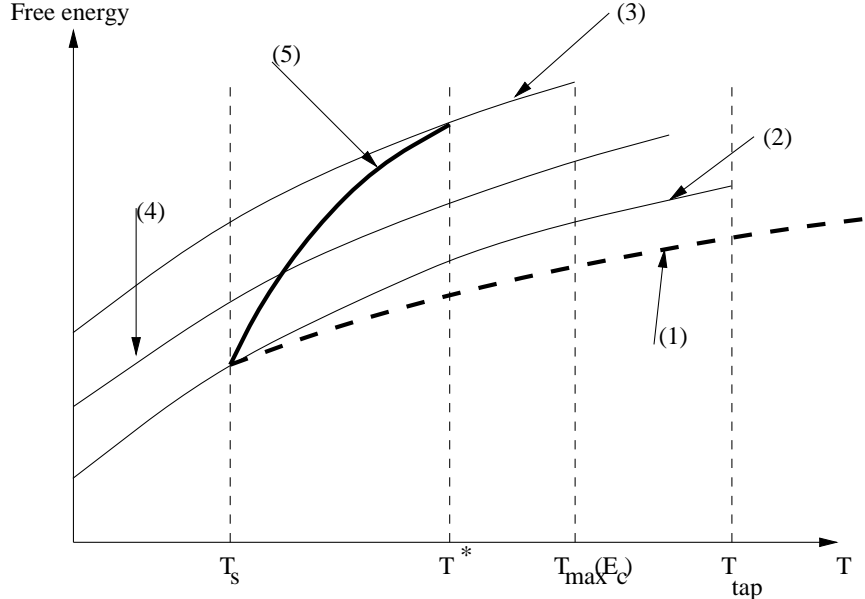,width=8cm,angle=-90}}}
\caption{Free energy versus temperature;
(1)~: free energy of the 
paramagnetic solution for $T > T^*$, $F_{tot}$ for $T < T^*$~; (2)~:
free energy of the lowest TAP states, with zero temperature
energy $E_{min}$; (3)~: free energy of the highest TAP states, corresponding
to $E_c$; (4)~: an intermediate value of $E_0$ leads to an 
intermediate value of $f$ at any temperature; (5)~: $f_{eq}(T)$; the
difference between curves (5) and (1) gives the complexity
$T S_c\((f_{eq}(T),T\))$.}
\label{figtap}
\end{figure}

The equation (\ref{ftap}) gives the free energy of a TAP state
with a given zero-temperature energy~;
to obtain the full partition function, we need to sum over the
possible energies, including the complexity term. After changing variables
we obtain an integral over the free energies of the TAP states
\cite{crisom95}~:
\be
Z= \int dF \exp \(( -{ (F-T S_c(F,T) ) \over T } \)).
\ee
For large $N$ we evaluate this integral by a saddle point method.
For $T > T^*$, with $T^* = \sqrt{p(p-2)^{p-2}(p-1)^{1-p}/2}$,
we find that the Boltzmann measure is dominated by the
paramagnetic state  $q=0$, with $F=N F_{para}$. On the contrary,
below $T^*$, this measure is dominated by TAP states
with free energy $F_{eq}$~; because of their complexity,
their total free energy is~:
\be
F_{tot} \equiv -T \ln (Z)= F_{eq}(T)-T S_c\((F_{eq}(T),T\)) \ .
\ee
For  $T < T_S$, these are the lowest TAP states, with $E^0 = E_{min}$,
and, for $T_S < T < T^*$, intermediate TAP states with a parameter 
$q$ given by~:
\be
\frac{p}{2T^2}q^{p-2}(1-q) = 1.
\label{qstat}
\ee
In this region $T_S < T < T^*$, $F_{tot}$ is equal to the
paramagnetic free energy. The global situation is pictured in 
f\/igure (\ref{figtap}).

Between the two  transition temperatures $T_s$ and $T_d$, the
situation is unclear~: the total equilibrium free energy seems
to get two equal contributions, from the paramagnetic state 
and from a bunch of TAP solutions
with non-zero $q$. One can wonder if there is a phase coexistence,
or simply a problem of double counting in the TAP approach. 
This issue, which is an important one
if one aims at understanding the finite dimensional behaviour
of this type
of systems \cite{parisi_verre}, can in fact be clarified within
a dynamical approach as. Let us also mention that some 
purely static approaches also carry relevant
information on related issues \cite{monasson,franzparisi}.
 
Before explaining this particular dynamical approach, we will recall
the previously known results of the equilibrium and
out of equilibrium dynamics.

\section{High temperature dynamics}

Following the study of the statics  \cite{crisom92},
Crisanti, Horner and Sommers studied the Langevin relaxation
dynamics of the model \cite{crihorsom}
\begin{equation}
\frac{ds_i(t)}{dt} = -\frac{\partial H}{\partial s_i} - \mu(t) s_i(t)
 + \eta_i(t),
\label{langepspin}
\end{equation}
where $\mu(t)$ has to be computed self consistently in order to 
implement the spherical constraint.
Using the same formalism as Sompolinsky and Zippelius for the SK model,
they wrote coupled equations for the correlation and response functions,
with random initial conditions corresponding to a quench at initial time.

\begin{figure}
\centerline{\hbox{
\epsfig{figure=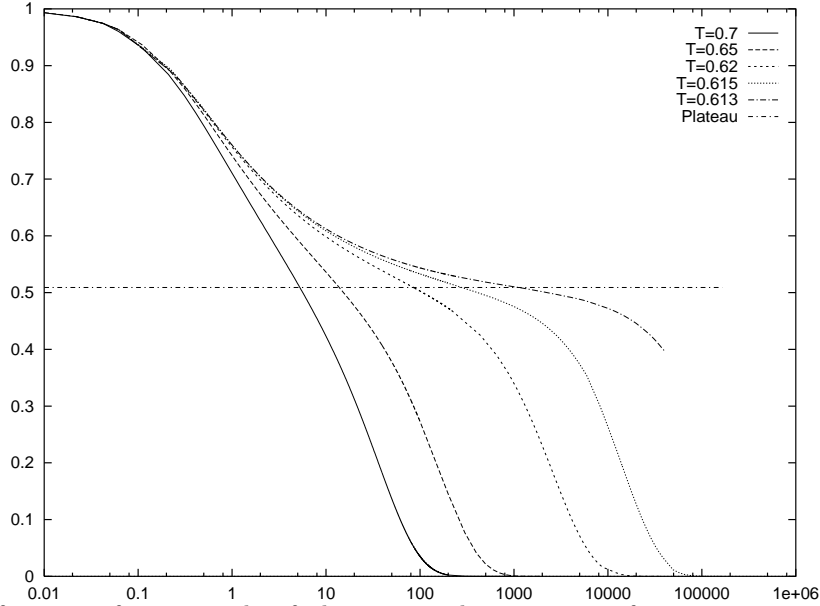,width=8cm,angle=-90}}}
\caption{$C_{eq}(\tau)$ as a function of $\tau$~: 
result of the numerical integration of equation 
equation (\ref{eqas}) for $p=3$, $T=0.7,\ 0.65,\ 0.62,\ 0.615,\
0.613$~; here $T_d=0.612372$. The horizontal line represents
the limit $\lim_{\tau \to \infty}C_{eq}(\tau)$ for $T=T_d$, here
$q \approx 0.509034$.}
\label{pspineq}
\end{figure}

If the validity of the fluctuation-dissipation theorem (FDT)
\be
\frac{\partial }{\partial t}C(t,t') = - T r(t,t'),
\ee
and invariance by time translation (TTI) are assumed,
the equations reduce to one single equation for the correlation
$C(t,t')=C_{eq}(t-t')$~:
\be
{\partial C_{eq}(\tau) \over \partial \tau} = - T C_{eq}(\tau)
- \frac{p}{2T} \int_0^{\tau}du\ C_{eq}^{p-1}(\tau-u)
\ \frac{\partial C_{eq}(u)}{\partial u},
\label{eqas}
\ee
with the initial condition $C_{eq}(0)=1$ (see appendix B.3.b).
These equation can be integrated numerically, with the following
result (f\/igure (\ref{pspineq}))~: at high temperature,
$\lim_{\tau \to \infty}C_{eq}(\tau)=0$~; if the temperature is lowered,
a plateau appears in the curve giving $C_{eq}(\tau)$ as a function
of $\log (\tau)$, which length diverges at a certain 
temperature $T_d$, given in appendix B.1.

Moreover, for $T > T_d$, with $T-T_d \ll T_d$, it can be shown analytically
\cite{crihorsom} that the plateau is approached with a power law,
$q + constant \ t^{-a}$, and the departure from the plateau
is in $q - constant' \ t^b$~; this behaviour is identical to the one
of the density correlation function in the mode coupling theories
concerning the glass transition (the analysis of Crisanti, Horner 
and Sommers \cite{crihorsom} follows the lines of the discussion
by G\"otze on the glass transition \cite{gotze2}).

The value of $T_d$ is precisely equal to the temperature
$T^*$ where appears the problem of double counting mentionned earlier.
Besides, the fact that $T_d$ is higher than the static transition
temperature shows that the transition at $T_d$ is purely dynamical.
Below $T_d$ the dynamics is no longer ergodic. and the validity of the
equilibrium dynamics properties should not be assumed.

\section{Out of equilibrium dynamics}

The dynamics below $T_d$ was studied in 
1993 by Cugliandolo and Kurchan \cite{cukuprl}, taking into account
the finite initial time. The dynamical equations for the correlation and
response functions are as follow (with $t>t'$, and random initial
conditions~; a possible derivation of this equations is
detailled in B.2)~:
\bea
\mu(t) &=& \int_0^t ds\  \frac{p^2}{2} C^{p-1}(t,s)
r(t,s) + T \nonumber \\
{\partial r(t,t') \over \partial t}&=& -\mu(t)  r(t,t')
+\frac{p(p-1)}{2} \int_{t'}^t ds  \ C^{p-2}(t,s)r(t,s)r(s,t')
\nonumber \\
{\partial C(t,t') \over \partial t} &=&-\mu(t)C(t,t')
+ \frac{p}{2} \int_0^{t'} ds \  C^{p-1}(t,s)r(t',s) \nonumber \\
&+& \frac{p(p-1)}{2} \int_0^t ds  \ C^{p-2}(t,s)r(t,s)C(s,t') .
\nonumber \\
\label{eqpspin}
\eea

\begin{figure}
\centerline{\hbox{
\epsfig{figure=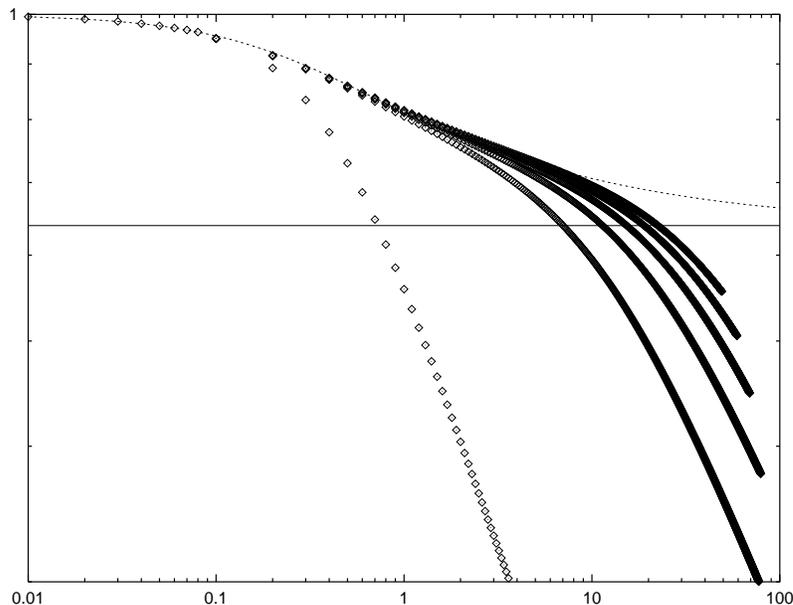,width=8cm,angle=-90}}}
\caption{p-spin model with $p=3$~:
$C(t_w+t,t_w)$ as a function of $t$ at temperature $T=0.5$
($T_d \approx 0.612$) for $t_w=0,\ 10,\ 20,\ 30,\ 40,\ 50$, in
logarithmic scale~; this curves were obtained by numerical integration
of the dynamical equations (\ref{eqpspin}), with the
algorithm used by [15]~; the
dotted curve is $C_{FDT}(t)$, the horizontal line corresponds
to $q$ at $T=0.5$, $q \approx 0.639$. 
}
\label{correlhorseq}
\end{figure}

At high temperature, the system (\ref{eqpspin}) reduces, via
TTI and FDT, to the dynamics studied by Crisanti,
Horner and Sommers, equation (\ref{eqas}), but, below
$T_d$, $C(t,t')$ and $r(t,t')$ depend on the two times.
The dynamical equations can be partially solved by the separation
of two time regimes \cite{cukuprl} (see f\/igure (\ref{correlhorseq}))~:
\begin{itemize}
\item FDT regime~:  for finite time separations $\tau=t-t'$,
i.e. for $\tau /t$ going to zero, the equilibrium dynamics
properties are valid, and we obtain two functions
$C_{FDT}(\tau)$ and $r_{FDT}(\tau)$ related by FDT. The
limit $\tau \to \infty$ yields $\lim_{\tau \to \infty} C_{FDT}(\tau)= q$,
with the value of the parameter of the threshold TAP states for $q$ 
(corresponding to  $E^0 = E_c$).

\item aging regime~: for $t$ and $t'$ going to infinity, without
$(t-t')/t \to 0$, i.e. for widely separated times, one can neglect
the time derivatives
${\partial r(t,t') \over \partial t}$,
${\partial C(t,t') \over \partial t}$, and this yields equations
invariant by time reparametrization. This way, an Ansatz can be found
to solve the equations~:
$C(t,t')= {\cal C}(h(t')/h(t))$, and
$r(t,t')= {\cal G}(h(t')/h(t)) h'(t')/h(t)$, with
${\cal C}(1)=q$, and $h$ a monotonously increasing function.
Moreover, ${\cal G}(\lambda) = x  {\cal C}'(\lambda)$,
with constant $x$. In this regime the correlation decreases from $q$
to zero.
\end{itemize}

The energy density can be shown to approach asymptotically
the energy density of the highest TAP states. As was noticed
in \cite{cukuprl}, this dynamics does not explore the TAP
states which dominate the Boltzmann distribution, and in fact
stays above all TAP states, with energy density going towards
the one of the threshold states, but staying at all times
$O(1)$ above.

\section{Inf\/luence of initial conditions}

Using the same field theoretical techniques as for the out of equilibrium
dynamics, it is possible to derive dynamical equations for a system
thermalized at temperature $T'$ at the initial time, and then
brought at temperature $T$.
However, to implement the Boltzmann measure at $T'$ for the initial
conditions, one has to reintroduce replicas (see appendix B.3, and
\cite{houghton,thirum1,thirum2,franzparisi}). The obtained equations
describe the evolution of two times overlaps between replicas
$C^{ab}(t,t')=\overline{<s^{a}(t)s^b(t')>}$, where $a$ and $b$ are replica
indices. These equations respect the initial structure (replica symmetric
or not) of the $C^{ab}$, i.e. the static replica structure describing
the equilibrium at temperature $T'$.

For the p-spin model, the initial situation is replica symmetric
if $T'$ is higher than the static transition temperature, with
$C^{ab}(0,0)=\delta_{ab}$. Therefore, at all times we can
write $C^{ab}(t,t')=C(t,t')\delta_{ab}$, and the
equations are now (for  $T'> T_s$, $t>t'$)~:
\bea
\mu(t) &=& \int_0^t ds\ \frac{p^2}{2} C^{p-1}(t,s)
r(t,s) + T + \frac{p}{2T'} C^{p}(t,0)  \nonumber \\
{\partial r(t,t') \over \partial t}&=& -\mu(t)  r(t,t')
+\frac{p(p-1)}{2} \int_{t'}^t ds  \ C^{p-2}(t,s)r(t,s)r(s,t')
\nonumber \\
{\partial C(t,t') \over \partial t} &=&-\mu(t)C(t,t')
+ \frac{p}{2} \int_0^{t'} ds \  C^{p-1}(t,s)r(t',s) \nonumber \\
&+& \frac{p(p-1)}{2} \int_0^t ds  \ C^{p-2}(t,s)r(t,s)C(s,t')
\nonumber \\
&+&
\frac{p}{2T'} C^{p-1}(t,0)\ C(t',0) .
\label{eqTT}
\eea

We see that the only difference with the usual dynamical equations
lie in terms of coupling to the initial condition. Of course, these
terms vanish in the limit  $T' \to \infty$.

\paragraph{$T=T'$}

If the temperature does not change at the initial time, $T=T'$, we
start with a thermalized system, and do not modify it. Therefore
the definition of the initial time is arbitrary and the system stays
in equilibrium~: the dynamics is equilibrium dynamics, with
$C(t,t')=C_{eq}(t-t')$, $r(t,t')=r_{eq}(t-t')$ and
$r_{eq}(\tau)=- \frac{1}{T} \frac{\partial C_{eq}}{\partial \tau}$.
The equations (\ref{eqTT}) yield then 
equation (\ref{eqas}) of equilibrium dynamics (see appendix
B.3).

For $T > T_d$ the system is in the paramagnetic state, with
$\lim_{\tau \to \infty} C_{eq}(\tau)=0$. On the other hand, below
$T_d$, we know that $C_{eq}(\tau)$ can not go to zero. We obtain
$\lim_{\tau \to \infty} C_{eq}(\tau)=C_\infty$, where $C_\infty$
is the highest solution of
\be
\frac{p}{2T^2}C_{\infty}^{p-2}(1-C_{\infty}) = 1
\ee
obtained taking $\tau \to \infty$ with
${\partial C_{eq}(\tau) \over \partial \tau} \to 0$ in the equation
of equilibrium dynamics (\ref{eqas}).
This equation is identical to (\ref{qstat}), which defines the
Edwards Anderson parameter of a TAP solution.

Therefore, below $T_d$, the thermalized system is in a TAP state,
and no more in a paramagnetic state~: we would then have
$C_{\infty}=0$. The Boltzmann measure is neither a superposition
of TAP states and a paramagnetic state, since the average on
initial composante would yield an intermediate value for
$C_{\infty}$.

This shows that, for temperatures lower than $T_d$,
the Boltzmann measure is given by the bunch of TAP states with
free energy $F_{eq}(T)$ such that
\be
F_{eq}(T)-T S_c\((F_{eq}(T),T\)) = F_{RS} = - \frac{1}{4T}.
\ee

\paragraph{$T$ dif\/ferent from $T'$}

\begin{figure}
\centerline{\hbox{
\epsfig{figure=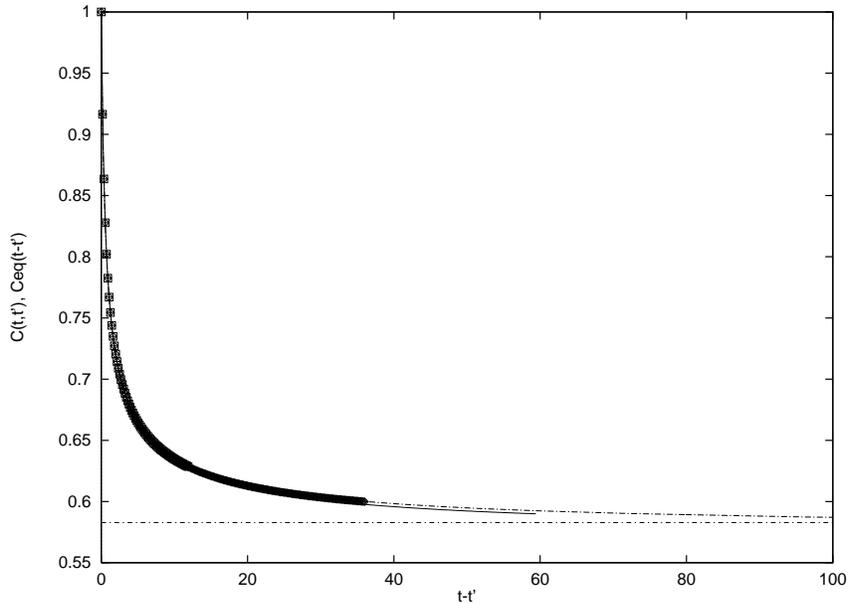,width=8cm,angle=-90}}}
\caption{$p=3$ model, with $T_S \approx 0.586$, $T_d \approx 0.612$~;
numerical integration of equations (\ref{eqTT}) for 
$T'=0.605$, $T=0.6$~; $C(t,0)$ is shown as a function of $t$ in full lines,
$C(t,t')$ as a function of $t-t'$ for $t'=6,\ 12,\ 18,\ 24$
(symbols)~; the dotted curve is the result of
the numerical integration of (\ref{as2}), and the horizontal line
represents the value of $C_\infty$ obtained by
(\ref{eqcinfini},\ref{eql}).
}
\label{pspineqtap}
\end{figure}

In order to push further the dynamical exploration of TAP states,
we can now study the dynamics at a temperature $T$ different from
the thermalization temperature $T'$. Two different scenarios are a priori
possible~: equations (\ref{eqTT}) can yield an explicit dependence on
$t$ and $t'$ for $C(t,t')$ and $r(t,t')$, or, after a transient regime, 
equilibrium dynamics, with functions of $t-t'$. A numerical integration
of the equations (using the algorithm of \cite{franzmezard,fm2})
leads to eliminate the first possibility (see figure
(\ref{pspineqtap})).

To study analytically the behaviour of the system, it is therefore
convenient to introduce as previously 
$C_{eq}(\tau)$, $r_{eq}(\tau)$, related by FDT, as well as
$C_\infty=\lim_{\tau \to \infty} C_{eq}(\tau)$,
$\mu_\infty=\lim_{t \to \infty}  \mu(t)$, and
$l=\lim_{t \to \infty} C(t,0)$. 
$l$ can differ from $C_\infty$ because of the transient regime.
We obtain~:
\bea
{\partial C_{eq}(\tau) \over \partial \tau}&=& - \left(\mu_{\infty} -
\frac{p}{2T} \right) C_{eq}(\tau) \nonumber \\
&+&\frac{p}{2} \int_0^{\tau}du\ C_{eq}^{p-1}(u)\ r_{eq}(\tau-u)
-\frac{p}{2T} C_{\infty}^p + \frac{p}{2T'} l^p.
\label{as2}
\eea

It is possible to extract the value of $C_{\infty}$~;
the energy of the system can also be computed \cite{cukuprl} (appendix B.3.b).

\begin{figure}
\centerline{\hbox{
\epsfig{figure=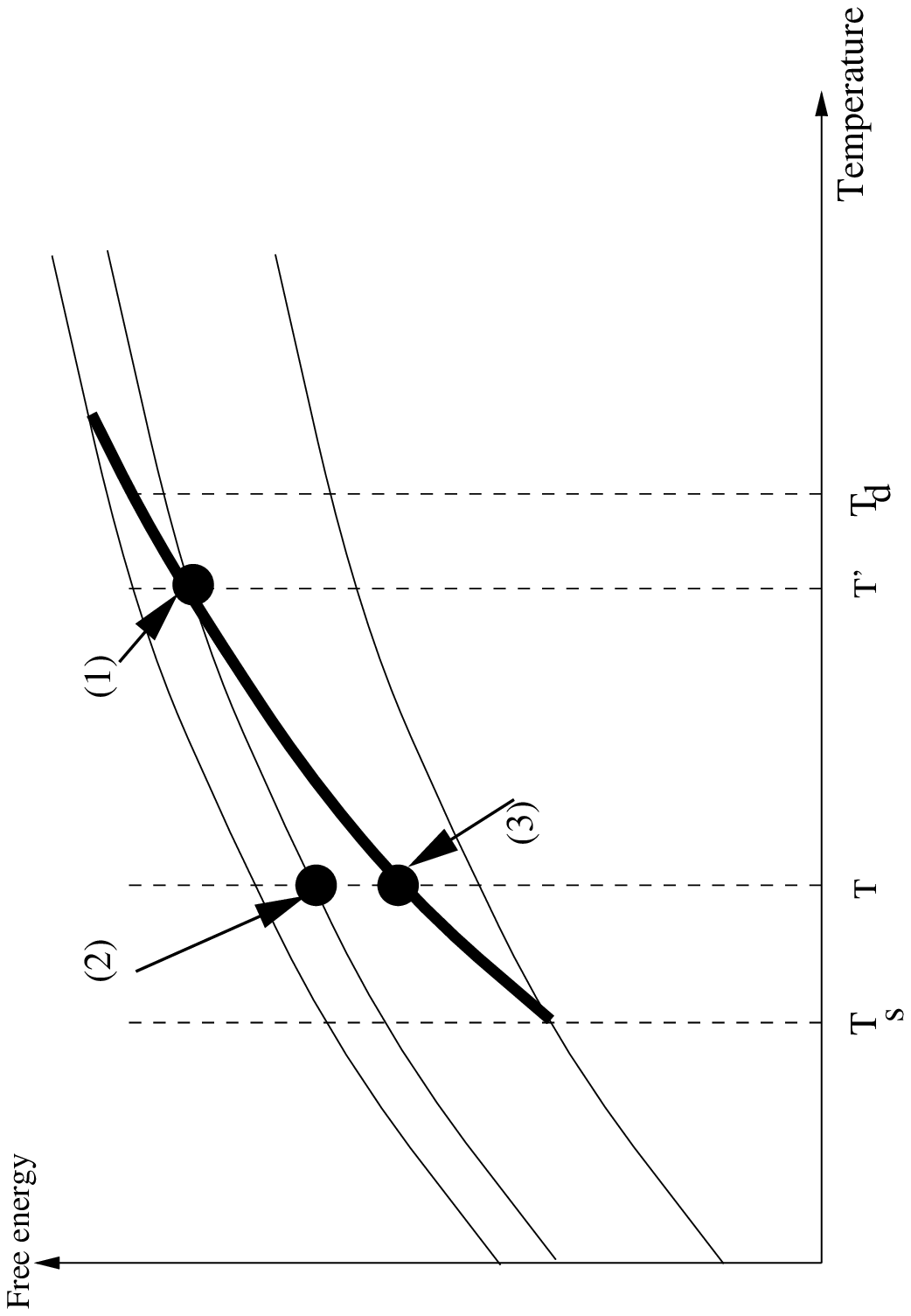,width=8cm,angle=-90}}}
\caption{
(1)~: TAP states giving the statics at $T'$, with angular energy
$E^{0}_{T'}$, and $q(E^{0}_{T'},T')$ ~;
(2)~: TAP states (1) followed at temperature $T$, with
$q(E^{0}_{T'},T)$ ~;
(3)~: TAP states giving the statics at $T$,  with angular energy
$E^{0}_{T}$, and $q(E^{0}_{},T)$.
}
\label{chgttap}
\end{figure}

For $T \neq T'$, we obtain that the asymptotic energy
$E_\infty$, and $C_{\infty}$ are precisely the energy and
the parameter $q$ of TAP states at temperature $T$~: those are
the TAP states obtained following the TAP states that are of
equilibrium at temperature $T'$ to the new temperature $T$~: these
states correspond to a certain angular energy $E^0_{T'}$, and $C_{\infty}$ is
equal to $q(E^0_{T'},T)$ (see f\/igure (\ref{chgttap})).

The obtained dynamics is therefore a relaxation in a TAP state,
in which the system was thermalized, followed at the new temperature  $T$.
It differs from the TAP states dominating the statics at $T$.
Moreover, taking the Laplace transform of (\ref{as2}), it is possible to 
show that the relaxation of $C_{eq}(\tau)$ for large  $\tau$ is of the
form $\tau^{-3/2} \exp (-\tau / \tau_0 )$. The 
relaxation  time $\tau_0$, which can be written as a function
of $T$, $T'$ and $C_\infty$ (the exact expression is complicated, and
I do not reproduce it here), diverges for the threshold TAP states
($E^0 = E_c$), and this relaxation occurs only for $T$ lower
than the temperature where the followed TAP state disappears.
If $T$ is further increased, we observe a fast relaxation in the 
paramagnetic state.

\section{Conclusions}

In this paper have been shown various aspects of
the spherical $p$-spin model, which, despite its relative
simplicity (it is a mean-field model, with a global spherical constraint
instead of a constraint for each spin) that allows for an
analytical treatment, displays a complex phase space and
interesting dynamical behaviours.

The phase space landscape is made of many metastable states
whose caracteristics are given by the ``TAP equations'', therefore
called ``TAP states''.

The use of particular initial conditions for the dynamics
has shown that
these solutions of the TAP equations correspond to real states,
with a real ergodicity breaking.
Below $T_d$, the equilibrium measure is given by TAP states, the paramagnetic
state vanishing at this dynamical transition.
The way in which this state disappears is still an open question, whose
answer would certainly help to understand the essence of the aging
behaviour obtained with a quench below $T_d$.

Moreover, if the system is prepared in a TAP state, by a thermalization
between $T_S$ and $T_d$, it stays trapped within, even if its 
temperature is changed~: these states can be followed by the dynamics
described in paragraph IV, as well for lower temperatures as
for higher ones, and even for temperatures higher than $T_d$,
until they disappear. They present a true ergodicity breaking even at 
these temperatures, but their complexity is not high enough to allow
them to contribute to the Boltzmann measure. 
The structure of these metastable states is still under investigation~:
in \cite{andreairenee} for example, it is shown that they are
not randomly distributed in the phase space.

On the other hand, the
usual dynamics at a temperature below $T_d$, starting
from a random configuration, only leads to
a ``weak ergodicity breaking'' \cite{bouchaud,cukuprl},
where the self overlap vanishes at very large time
differences (much larger than the waiting time).
This is explained\cite{cukuprl,kurchanlaloux}
by the fact that the system, which was initially
in the (infinite temperature) paramagnetic state, does
not find any TAP state in a finite time, but stays
at energy density $O(1)$ (going to zero as $t$ goes to infinity)
above the threshold.

It is likely that the impossibility for the system to find
the states in a finite time results from the mean-field approximation,
and that, in finite dimensions, the time to find these states would be
finite, the dynamics being obtained afterwards by jumps between states.
Recent numerical simulations of a model with $4$ spins-interactions in
three dimensions
\cite{franzpspin} tend to support this scenario.

{\bf Acknowledgments}

It is a pleasure to thank my
PhD thesis advisor M. M\'ezard for useful discussions and comments~;
the paragraph IV contains results
obtained in collaboration with R. Burioni and M. M\'ezard.
\cite{babumez}.

\newpage

\appendix

\section{The statics}

\subsection{Replicas}

The study of the $p$-spin model with the replica method yields
the following results~:
\begin{itemize}
\item  at high temperature, the system is replica symmetric, with
$Q_{EA}=0$ (paramagnetic region)~;
\item below the static transition temperature
\be
T_S = y \sqrt{ \frac{p}{2y} } (1-y)^{\frac{p}{2} -1} \ , \mbox{where}
\ \ \frac{2}{p} = -2y \frac{1-y +\ln y }{(1-y)^2}\ ,
\ee
there is a one-step replica symmetry breaking, with parameters
$q_0 =0$, $q_1$, $x$, and free energy density
\be
F_{RSB} = -\frac{1}{4T} \(( 1 - (1-x) q_1^p \))
+ \frac{T}{2x} \ln \(( \frac{1-q_1}{1 - (1-x)q_1} \))
-\frac{T}{2} \ln (1-q_1)
\ee
and
\be
q_1^{\frac{p}{2} -1} (1-q_1) = \sqrt{\frac{2y}{p}} T \ ,
\ x= T q_1^{-\frac{p}{2}} \sqrt{\frac{2y}{p}} \(( \frac{1}{y} -1 \)).
\ee
\end{itemize}

\subsection{TAP solutions}

\subsubsection{Derivation of the TAP equations}
 
We derive here the TAP equations on the local magnetizations, using 
the cavity method \cite{mezparvir}.
 
The Hamiltonian is
\be
H =  - \sum_{1 \leq i_1 < i_2 \cdots < i_p \leq N}
J_{i_1 i_2 \cdots i_p} s_{i_1}s_{i_2}\cdots s_{i_p}
- \frac{r}{2} \sum_{i=1}^{N} \si^2
\ee
where $r$ is ajusted in a self consistent way to implement the
spherical constraint.
 
To the Hamiltonian for $N$ spins $i=1, \cdots, N$ we add a spin
$s_0$, with coupling constants $J_{0 i_2\cdots i_p}$.
The field acting on $s_0$ is
\be
h = \sum_{1 \leq  i_2 < \cdots < i_p \leq N}
J_{0 i_2\cdots i_p}s_{i_2}\cdots s_{i_p}.
\ee
In the absence of $s_0$, to each configuration of the
$\si$ correspond a value of $h$. The probability distribution of this
field in the absence of $s_0$ will allow us to obtain the joint
distribution of $s_0$ and $h$.
The mean value of $h$ is~:
\be
\langle h \rangle_{ _N} = \sum_{1 \leq  i_2 < \cdots < i_p \leq N}
J_{0 i_2\cdots i_p} \langle  s_{i_2}\rangle_{ _N}  \cdots
\langle  s_{i_p}\rangle_{ _N}
\ee
since the $J_{0 i_2\cdots i_p}$ and the $\si$ are independant variables~;
we have also used the hypothesis
$\langle  s_{i_2} \cdots s_{i_p}\rangle_{ _N}
\approx \langle s_{i_2} \rangle_{ _N} \cdots \langle  s_{i_p}
\rangle_{ _N}$, so called ``clustering'' hypothesis \cite{mezparvir},
more precisely
\be
\lim_{N \to \infty} \frac{1}{N^{p-1}}
\sum_{i_2 \cdots i_p} \(( \langle s_{i_2} \cdots s_{i_p}\rangle_{ _N}
- \langle s_{i_2} \rangle_{ _N} \cdots \langle  s_{i_p} \rangle_{ _N}
\))^2 = 0.
\ee
Besides, in
\bea
\langle (h - \langle h \rangle_{ _N})^2 \rangle_{ _N} & &=
\sum_{ 1 \leq  i_2 < \cdots < i_p \leq N , 
1 \leq  k_2 < \cdots < k_p \leq N }
J_{0 i_2\cdots i_p} J_{0 k_2\cdots k_p} \\
& &\langle (s_{i_2}\cdots s_{i_p} - \langle  s_{i_2}\rangle_{ _N}  \cdots
\langle  s_{i_p}\rangle_{ _N}) 
(s_{k_2}\cdots s_{k_p} - \langle  s_{k_2}\rangle_{ _N}  \cdots
\langle  s_{k_p}\rangle_{ _N}) \rangle
\eea
the terms such that $\{i_2, \cdots , i_p\}=\{k_2, \cdots , k_p\}$
dominate, and
\be
\langle (h - \langle h \rangle_{ _N})^2 \rangle_{ _N} =
\frac{p}{2} ( 1 - q^{p-1}).
\ee
It is possible to show (calculating higher moments) to show that the
distribution of $h$ in the absence of $s_0$ is gaussian.
 
In the rpesence of $s_0$, we add an interaction $h s_0$, so the joint
distribution of $h$ and $s_0$ is of the form
\be
P_{N+1} (h,s_0) \propto
\exp \(( - \frac{\beta r}{2} s_0^2 + \beta h s_0 
- \frac{(h-\langle h \rangle_{ _N})^2}{p ( 1 - q^{p-1} )} \)).
\ee
We deduce
\be
\langle s_0 \rangle_{ _{N+1}} = \frac{\langle h \rangle_{ _N}}
{r - \frac{p\beta}{2}(1 - q^{p-1} )}
\ee
and
\be
\langle ( s_0 - \langle s_0 \rangle_{ _{N+1}})^2 \rangle_{ _{N+1}}
= \frac{1}
{\beta(r - \frac{p\beta}{2}(1 - q^{p-1} ))} .
\ee
Because of the spherical constraint, 
$\langle ( s_0 - \langle s_0 \rangle_{ _{N+1}})^2 \rangle_{ _{N+1}}
=1 - q$, so that~:
{\boldmath\be
\colorbox{grey0.95}{\makebox{
$ \displaystyle
\beta r =  \frac{1}{1-q} + \frac{p \beta^2}{2}(1- q^{p-1}).
$
}}
\label{rq}
\ee}
 
We need equations between the
$\langle s_i \rangle_{ _{N+1}}$, therefore we must obtain
$\langle h \rangle_{ _N}$ as a function of these magnetizations.
In order to do this,
we look at the influence on $\langle s_i \rangle$ of the new spin
$s_0$, for a given $i$.
We write
\be
h = s_i h_i + h_i'
\ee
where
\be
h_i = \sum_{1 \leq  i_3 < \cdots < i_p \leq N}
J_{0 i i_3\cdots i_p}  s_{i_3} \cdots s_{i_p} ,
\ee
the sum being on indices non equal to $i$.
Then $h_i$ is of order $N^{-1/2}$, and to leading order
\bea
\langle (h_i - \langle h_i \rangle_{ _N})^2 \rangle_{ _N} &=&
\frac{p(p-1)}{2N} (1 - q^{p-2}) \\
\langle (h_i' - \langle h_i' \rangle_{ _N})^2 \rangle_{ _N} &=&
\frac{p}{2} (1 - q^{p-1})
\eea
so the joint probability distribution of $s_0$, $s_i$,
$h_i$, $h_i'$ is
\bea
P_{N+1}(s_0,s_i,h_i,h_i') \propto \exp \((
- N \frac{(h_i-\langle h_i \rangle_{ _N})^2}{p(p-1)( 1 - q^{p-1} )} 
- \frac{(h_i' -\langle h_i' \rangle_{ _N})^2}{p ( 1 - q^{p-1} )} 
\right. \\
\left. - \frac{\beta r}{2} s_0^2 + \beta s_0 (s_i h_i + h_i')
- \frac{(s_i -\langle s_i \rangle_{ _N})^2}{2(1-q)} \)).
\eea
The term $\frac{(s_i -\langle s_i \rangle_{ _N})^2}{2(1-q)}$ comes
from the fact that, in the absence of  $s_0$, the spin $s_i$ has
average 
$\langle s_i \rangle_{ _N}$ and variance $1-q$.
Integrating out $h_i$, $h_i'$ yields
\be 
P_{N+1}(s_0,s_i) \propto \exp \((
- \frac{\beta r}{2} s_0^2 + \frac{p \beta^2}{4}(1-q^{p-1})s_0^2
+ \beta s_0 (s_i \langle h_i \rangle_{ _N} +
\langle h_i' \rangle_{ _N} )
- \frac{(s_i -\langle s_i \rangle_{ _N})^2}{2(1-q)} \)),
\ee
anf finally with (\ref{rq})~:
\be
\langle s_i \rangle_{ _{N+1}} =
\langle s_i \rangle_{ _N} + \beta^2 (1-q)^2 
\langle h_i \rangle_{ _N} \langle h_i' \rangle_{ _N}
\ee
or
\be
\langle s_i \rangle_{ _{N+1}} =
\langle s_i \rangle_{ _N} + \beta (1-q)
\langle s_0 \rangle_{ _{N+1}} 
\sum_{1 \leq  i_3 < \cdots < i_p \leq N}
J_{0 i i_3\cdots i_p}  \langle s_{i_3} \rangle_{ _N}
 \cdots \langle  s_{i_p}\rangle_{ _N}.
\ee
The technique used by Rieger \cite{riegertap} allows to write
the TAP equations on
$m_i= \langle s_i \rangle_{ _{N+1}}$, $i=0, \cdots , N$~:
as
\be
\frac{m_0}{\beta(1-q)} =
\sum_{1 \leq  i_2 < \cdots < i_p \leq N}
J_{0 i_2\cdots i_p} \langle  s_{i_2}\rangle_{ _N}  \cdots
\langle  s_{i_p}\rangle_{ _N} ,
\ee
we obtain
\be
\frac{m_0}{\beta(1-q)} =
\sum_{1 \leq  i_2 < \cdots < i_p \leq N}
J_{0 i_2\cdots i_p} m_{i_2} \cdots m_{i_p} - A ,
\ee
where
\be
A=
(p-1) \beta(1-q) m_0 \sum_{1 \leq  i_2 < \cdots < i_p \leq N}
J_{0 i_2\cdots i_p} m_{i_2} \cdots m_{i_p}
\sum_{1 \leq  k_3 < \cdots < k_p \leq N}
J_{0 i_2 k_3\cdots k_p} m_{k_3} \cdots m_{k_p} , 
\ee
and, using the coupling symmetry under permutations~:
\bea
A &=& (p-1) \beta(1-q) m_0 \frac{1}{(p-1)!}
\sum_{i_2  \cdots  i_p}
J_{0 i_2\cdots i_p} m_{i_2} \cdots m_{i_p} \frac{1}{(p-2)!}
\sum_{k_3 \cdots k_p} J_{0 i_2 k_3\cdots k_p} m_{k_3} \cdots m_{k_p} \\
&=& \beta(1-q) m_0 \frac{1}{(p-2)!^2}
\sum_{i_2} \(( \sum_{k_3 \cdots k_p} J_{0 i_2 k_3\cdots k_p} m_{k_3}
\cdots m_{k_p} \))^2 .
\eea
The terms $J_{0 i_2 k_3\cdots k_p}^2 m_{k_3}^2 \cdots m_{k_p}^2$
are dominant and yield~:
\be
A = \beta(1-q) m_0 \frac{1}{(p-2)!^2} \sum_{i_2} (p-2)!
\frac{p!}{2 N^{p-1}} (Nq)^{p-2}
\ee
The equations are finally~:
{\boldmath\be
\colorbox{grey0.95}{\makebox{
$ \displaystyle
\frac{m_0}{\beta(1-q)} = 
\sum_{1 \leq  i_2  < \cdots < i_p \leq N} J_{0 i_2 \cdots i_p}
m_{i_2} \cdots m_{i_p} 
- \beta (1-q) \frac{p(p-1)}{2} q^{p-2} m_0.
$}}
\ee}

\subsubsection{Study of the TAP states}
 
The states $\alpha$ are given by
\be
m_i^\alpha= \sqrt{q (E^0_\alpha,T )} \hat{s}_i^\alpha \ ,
\ee
where $q(E,T)$ is obtained by minimizing the free energy
\be
q^{\frac{p}{2}} E - \frac{T}{2}\ln(1-q) -
\frac{1}{4T}[(p-1)q^p - pq^{p-1} +1] \ ~;
\label{el}
\ee
if we take
\be
z = \frac{1}{T} (1-q) q^{\frac{p}{2}-1}
\ee
we obtain the equation
\be
\frac{p(p-1)}{2} z^2 + p z E + 1 =0 ,
\ee
which yields that $q$ is the largest solution of
{\boldmath
\be
\colorbox{grey0.95}{\makebox{
$ \displaystyle
(1-q) q^{\frac{p}{2}-1} = T \(( -E-\sqrt{E^2-E_c^2} \over p-1 \)) \ ,
$
}}
\label{qtap}
\ee}
with $E_c= -\sqrt{2(p-1)/p}$ (the other solution is not a
minimum of the free energie (\ref{el})).
 
The maximum of $(1-q) q^{p/2-1}$ occurs for $q=1 - \frac{2}{p}$,
so that the equation (\ref{qtap}) has no more solutions when
\be
\frac{2}{p}  \(( 1- \frac{2}{p}\))^{\frac{p}{2} -1} = 
T \(( -E-\sqrt{E^2-E_c^2} \over p-1 \)) \ .
\ee
Therefore, a TAP state of energy  $E^0$ disappears at
\be
T_{max}(E^0) = \(( \frac{2}{p} \))
\(( \frac{p-1}{ -E^0 -\sqrt{(E^0)^2-E_c^2}}  \))
 \((1- \frac{2}{p}\))^{\frac{p}{2} -1} .
\ee
 
The energy per spin of a state $\alpha$, given by
$\frac{\partial (\beta f)}{\partial \beta}$, is~:
\be
E_\alpha=q_\alpha^{\frac{p}{2}} E^0_\alpha - \frac{1}{2T}
[(p-1)q_\alpha^p - pq_\alpha^{p-1} +1] \ .
\label{etap}
\ee
 
The number of TAP states with angular energy $E^0$ is exponential
in $N$, and the complexity is \cite{crisom95}~:
{\boldmath
\be
\colorbox{grey0.95}{\makebox{$\displaystyle
s_c^0(E^0) = 
 \frac{1}{2} \(( - \ln \frac{p z^2}{2} + \frac{p-1}{2}z^2
-\frac{2}{p^2 z^2} + \frac{2-p}{p} \)) ,
$}}
\ee
}
with $z= \((-E^0-\sqrt{ (E^0)^2-E_c^2}\))/(p-1)$.
This complexity goes to zero for
\be
z = \sqrt{\frac{2y}{p}},
\ee
which corresponds to $E^0 = E_{min}=E_{RSB}(T=0)$.
Therefore there exist TAP states for zero teperature energies
between $E_{min}$ and $E_c$.
 
The lowest states disappear at
{\boldmath
\be
\colorbox{grey0.95}{\makebox{$\displaystyle
T_{TAP} = \frac{2}{p} \sqrt{\frac{p}{2y}} \(( 1- \frac{2}{p}
\))^{\frac{p}{2} -1} .
$}}
\ee
}

Let us now determine the parameters of the TAP states dominating
the partition function. They minimize
\be
F_{tot} \equiv -T \ln (Z)= F_{eq}(T)-T S_c\((F_{eq}(T),T\)) \ ,
\ee
so that
\bea
& &q^{\frac{p}{2}} E^0 - \frac{T}{2}\ln(1-q) -
\frac{1}{4T}[(p-1)q^p - pq^{p-1} +1] \\
& &- \frac{T}{2} \(( - \ln \frac{p z^2}{2} + \frac{p-1}{2}z^2
-\frac{2}{p^2 z^2} + \frac{2-p}{p} \)) .
\eea
Taking the derivative with respect to $q$, using the relation
$\frac{p(p-1)}{2} z^2 + p z E^0 + 1 =0$, substituing
$z$ with its expression as a function of $q$, and taking
\be
x = \frac{p}{2T^2} (1-q) q^{p-2},
\ee
we obtain
\be
(1-x)((p-1)(1-q)x -1)=0,
\ee
which yields~:
{\boldmath
\be
\colorbox{grey0.95}{\makebox{$\displaystyle
\frac{p}{2T^2} (1-q) q^{p-2}=1 
$}}
\label{qtapeq}
\ee}
or
\be
\frac{p(p-1)}{2T^2} (1-q)^2q^{p-2}=1.
\ee
 
The first solution yields a free energy density equal to 
$-1/(4T)$, i.e.. equal to the free energy density of the paramagnetic
solution. Moreover~:
\bea
z &=& \sqrt{\frac{2(1-q)}{p}} \\
E^0 &=& \sqrt{\frac{p}{2(1-q)}} \(( \frac{p-1}{p} q -1 \)).
\label{en0tap}
\eea
Given that $(1-q) q^{p-2}$ is maximum for $q=(p-2)/(p-1)$, this solution
is only possible for temperatures lower than $T^*$, with 
{\boldmath
\be
\colorbox{grey0.95}{\makebox{$\displaystyle
T^* = \sqrt{ \frac{p(p-2)^{p-2}}{2(p-1)^{p-1}} } .
$}}
\ee}
Above $T^*$, the paramagnetic solution with $q=0$ is therefore dominant.
For $T=T_S$, $q$ goes to the value $q_1$ of the lowest TAP states
Below $T_S$ there is no more any solution, and these states, with
the lowest total free energy, dominate.

The second solution corresponds to the threshold states, and is valid
only for $T=T^*$~; for any other value of $T$ it does not yield the
lowest free energy.
 
\newpage
 
\section{Dynamics of the p-spin spherical model}
\subsection{Dynamical transition}
 
The dynamical transition temperature can be obtained by the dynamical
stability criterion~: the correlation function is a decreasing
function of the time${\partial C_{eq}(\tau) \over \partial \tau} \le 0$.
This criterion, together with the evolution equation of
$C_{eq}$ (\ref{eqas}), commands that, at all times,
\be
 - T C_{eq}(\tau) - \frac{p}{2T} C_{eq}^{p-1}(\tau) (C_{eq}(\tau)-1)
\leq  0.
\ee
It is now easy to check that this inequality cannot be satisfied
for all times, if
$\lim_{\tau \to \infty}C_{eq}(\tau)=0$, below $T_d$, with
{\boldmath
\be
\colorbox{grey0.95}{\makebox{$\displaystyle
 T_d = \sqrt{\frac{p (p-2)^{p-2}}{2 (p-1)^{p-1}}} .
$}}
\ee}
 
\subsection{Dynamical equations}
 
The starting point of the formalism is the equality
\be
1 = \int {\cal D}s_1 \cdots {\cal D}s_N
\prod_t \prod_{i=1}^N \delta \(( 
 dt \(( \sid(t) + \frac{\partial H}{\partial
\si} + \mu(t) \si(t) - \eta_i(t) \)) \)),
\ee
which is simply a rewriting of the Langevin equation
(the corresponding Jacobian is $1$ if the It\^{o} discretization
scheme is used for the Langevin equation, i.e. if
$\si(t+ \delta t)=\si(t) - \delta t \frac{\partial H}{\partial \si}
(t) -\delta t \mu(t) \si(t) + \int_t^{t+\delta t} \eta_i(t') dt'$).
 
The notation ${\cal D}s_i$ stands for 
$\prod_t ds_i(t)$~: at each time $t$ the integration is performed
over all possible values of the spins.
 
Writing  ${\cal D}s {\cal D} \hat{s}$ for
${\cal D}s_1 \cdots {\cal D}s_N{\cal D} \hat{s}_1 \cdots {\cal D}
\hat{s}_N$ we obtain~:
\be
1 = \int {\cal D}s {\cal D} \hat{s}
\exp \(( \sum_{i=1}^N \int_0^{\infty}
 dt \sih(t) \(( \sid(t) +
\frac{\partial H}{\partial \si} + \mu(t) \si(t) - \eta_i(t) \)) \)).
\ee
 
This equality can be averaged over the noise~:
\be
1 = \int {\cal D}s {\cal D} \hat{s}
\exp \(( \sum_{i=1}^N \int_0^{\infty}
 dt \(( 
T \sih^2(t) + \sih(t) \sid(t) + \sih(t) \frac{\partial H}{\partial \si}
 + \mu(t) \sih(t) \si(t) \)) \)),
\ee
and then over the quenched disorder, using
\be
\frac{\partial H}{\partial \si} =
- \((\sum_{i < i_2 \cdots < i_p} J_{i i_2 \cdots i_p}
+ \sum_{i_2 < i < i_3 \cdots < i_p} J_{i_2 i i_3 \cdots i_p} +
\cdots
+ \sum_{i_2 < \cdots < i_p< i } J_{i_2  i_3 \cdots i_p i} \))
s_{i_2}\cdots s_{i_p}~;
\label{dhdsi}
\ee
we note
\be
\sum_i^* = \sum_{i < i_2 \cdots < i_p}
+ \sum_{i_2 < i < i_3 \cdots < i_p} +
\cdots + \sum_{i_2 < \cdots < i_p< i },
\ee
and similarly $\sum_{ij}^*$ the sum $\sum_{i_3 \cdots < i_p}$
with indices $i_k$ dif\/ferent from $i$ and from $j$~:
\bea
1 &=& \int {\cal D}s {\cal D} \hat{s} \exp S \\ \nonumber 
S &=& \sum_{i=1}^N \int_0^{\infty}
 dt \((T \sih^2(t)+ \sih(t) \sid(t)
+ \mu(t) \sih(t) \si(t) \)) \\ \nonumber
&+& \frac{p!}{4 N^{p-1}} \(( \sum_{i=1}^N \sum_i^* \int_0^{\infty} dt 
\int_0^{\infty} dt'
\sih(t)\sih(t') s_{i_2}(t)s_{i_2}(t') \cdots s_{i_p}(t)s_{i_p}(t')
\right.
 \\ 
&+& \left. \sum_{i,j=1~; i \neq j}^N \sum_{ij}^* 
\int_0^{\infty} dt \int_0^{\infty} dt'
\sih(t)\si(t') s_j(t) \hat{s}_j(t') s_{i_3}(t)s_{i_3}(t') \cdots
s_{i_p}(t)s_{i_p}(t') \)) .
\label{action}
\eea
 
By definition of the correlation and response functions, these
quantities are given by the average of
$\si(t) \si(t')$ and $\si(t) \sih(t')$ with the action $S$.
Taking the limit $N \to \infty$, we will therefore be able to write
in a self consistent way~:
\be
\frac{1}{N} \sum_i \si(t) \si(t') = C(t,t')~; \ \ 
 - \frac{1}{N} \sum_i \si(t) \sih(t') = r(t,t'),
\ee
which is equivalent to using the saddle point method. Besides,
for $N \to \infty$, $\frac{1}{N} \sum_i \sih(t) \sih(t') =0$
(it is a double derivative of $1$ with respect to
$\eta_i(t)$ and $\eta_i(t')$).

We then use~:
\be
\langle A \frac{\partial S}{\partial \sih(t)} \rangle
= \langle \frac{\partial A}{\partial \sih(t)} \rangle,
\ee
with  $A = \si(t')$ and $A = \sih(t')$~; 
\bea
\frac{\partial S}{\partial \sih(t)} &=& 
2 T \sih(t)+ \sid(t) + \mu(t) \si(t) \\ \nonumber
&+& \frac{p!}{2 N^{p-1}} \int_0^{\infty} dt' \sih(t') \sum_i^*
s_{i_2}(t)s_{i_2}(t') \cdots s_{i_p}(t)s_{i_p}(t') \\
&+& \frac{p!}{2 N^{p-1}} \int_0^{\infty} dt' \si(t') \sum_{j=1~; j \neq i}^N
s_j(t) \hat{s}_j(t') \sum_{ij}^* s_{i_3}(t)s_{i_3}(t') \cdots 
s_{i_p}(t)s_{i_p}(t').
\eea
 
We multiply by $\si(t')$, and sum over $i$~; in the limit
$N \to \infty$ we obtain~:
\bea
0 =&-& 2 T r(t',t) + \frac{\partial C(t,t')}{\partial t}
+ \mu(t) C(t,t') \\ \nonumber
&-& \frac{p}{2} \int_0^{t'} dt'' r(t',t'') C(t,t'')^{p-1} \\
&-& \frac{p(p-1)}{2} \int_0^t dt'' C(t',t'')r(t,t'')C(t,t'')^{p-2}
\eea
(with $r(t',t)=0$ for $t>t'$).
 
In the same way, multiplying by $\sih(t')$ and summing over $i$
yields the equation for $r$.
Taking the limit $t' \to  t$, with $C(t,t)=1$, yields then the
self consistency equation for $\mu(t)$.
 
The energy is obtained multiplying 
(\ref{langepspin}) by $s_i(t')$, averaging over the noise $\eta_i$
and the couplings, and finally taking the limit $t' \to t$~:
\be
\sum_i \si(t') \sid(t) = -\sum_i s_i(t')
\frac{\partial H}{\partial \si}(t)  -\mu(t) \sum_i s_i(t') \si(t)
+ \sum_i \si(t')\eta_i(t)  \\ \nonumber
\ee
and, with (\ref{dhdsi})~:
\be
\overline{ \frac{1}{N}
\sum_i \si(t) \frac{\partial H}{\partial \si}(t)
}= p E(t)
\ee
so, with $\lim_{t' \to t} \partial_t C(t,t') = - T$~:
{\boldmath
\be
\colorbox{grey0.95}{\makebox{$\displaystyle
E(t)= \frac{ T - \mu(t) }{p}.
\label{energie}
$}}
\ee}

\subsection{Thermalized initial conditions}
 
\subsubsection{Dynamical equations}
 
In the previous derivation, the average over the
initial conditions $\si(0)$ was not performed, since the
system was taken disordered at $t=0$. This average is therefore
uniform.
 
If on the other hand the initial system is thermalized at a given
temperature $T'$, this average is
\be
\int {\cal D} s(0) \cdots \frac{ e^{-H[\si(0)]/T'}}{ Z(T')},
\ee
where $Z(T')$ is the partition function at $T'$. As we want to average
over the couplings, we have to reintroduce replicas~:
\be
\frac{1}{ Z(T')}= \lim_{n \to 0}
\int {\cal D} s^1 \cdots {\cal D} s^{n-1} \exp \((
- \frac{1}{T'} \sum_{a=1}^{n-1} H[\si^a] \)).
\ee
 
We introduce at all times $n$ replicas, with the same disorder
realization. The spins are now
$\si^a(t)$ with $i=1,\cdots,N$, $a=1,\cdots,n$.
We obtain after averaging over the noise
\bea
1 &=& \int {\cal D}s^a {\cal D} \hat{s}^a
\exp  \sum_{a=1}^n  S_J^a \\
S_J^a &=& \sum_{i=1}^N \int_0^{\infty} dt \sih^a(t)
\(( T \sih^a(t) + \sid^a(t) + 
\frac{\partial H^a}{\partial \si^a} + \mu(t) \si^a(t)
\)) 
- \frac{1}{T'} H[\si^a(0)] .
\eea
 
After the average over the couplings, we have the same terms as in
(\ref{action}), with a sum over replica indices, and the
following new terms~:
\be
\frac{1}{T'} \frac{p!}{4 N^{p-1}}
\sum_{a \leq b} \sum_i \sum_i^* \int_0^{\infty} dt 
\sih^a(t) \si^b(0) s_{i_2}^a(t)s_{i_2}^b(0) \cdots
s_{i_p}^a(t)s_{i_p}^b(0).
\label{termessup}
\ee
 
Taking the derivative with respect to $\sih^a(t)$, multiplying
by  $\si^b(t')$ or  $\sih^b(t')$, and summing over $i$, we
obtain for $N \to \infty$ equations for  $C^{ab}(t,t')$ and $r^{ab}(t,t')$,
with $r^{ab}(t,t')=\delta_{ab}r(t,t')$  since the replicas are
not coupled. The terms (\ref{termessup}) give rise to
a coupling to the initial conditions
$C^{ab}(t,0)$ and $C^{ab}(t',0)$. We thus obtain
equations which, in the RS case, reduce to (\ref{eqTT}).
 
\subsubsection{Case T=T'}
 
The system is FDT and TTI~: $C(t,t')= C_{eq}(t-t')$,
$r(t,t')= r_{eq}(t-t')$~; we note $\tau = t-t'$~; then
$r_{eq}(\tau)=- \frac{1}{T} \frac{\partial C_{eq}}{\partial \tau}$.
 
The integral appearing in the equation for $\mu(t)$ is~:
\be 
\frac{p^2}{2}\int_0^t ds\ C^{p-1}(t,s)r(t,s) = - \frac{1}{T}
\frac{p^2}{2} \int_0^t du\ C_{eq}^{p-1}(u)C_{eq}'(u) =
- \frac{p}{2T} (C_{eq}^p(t) - 1) 
\ee
so that
\be
\mu(t) = T - \frac{p}{2T} (C_{eq}^p(t) - 1) +\frac{p}{2T} C_{eq}^p(t)
= T+\frac{p}{2T} .
\ee
 
For the equation for the correlation, the first integral is
\be 
\int_0^{t'} ds \  C^{p-1}(t,s)r(t',s) = 
\int_0^{t'} ds \  C_{eq}^{p-1}(t-s) r_{eq}(t'-s)  
=  - \frac{1}{T} \int_\tau^{t} du \  C_{eq}^{p-1}(u)  C_{eq}'(u-\tau),
\ee
and the second one yields
\bea
(p-1)\int_0^t ds  \ C^{p-2}(t,s)r(t,s)C(s,t') &=& 
(p-1)\int_0^{t'} ds  \ C_{eq}^{p-2}(t-s)r_{eq}(t-s)C_{eq}(t'-s) 
\\ \nonumber
&+& (p-1)\int_{t'}^t ds  \ C_{eq}^{p-2}(t-s)r_{eq}(t-s)C_{eq}(s-t')
\\ \nonumber
&=& - \frac{p-1}{T} \int_\tau^t du \ C_{eq}^{p-2}(u)C_{eq}'(u)
C_{eq}(u-\tau) \\ \nonumber
&-& \frac{p-1}{T} \int_0^\tau du \
C_{eq}^{p-2}(u)C_{eq}'(u)C_{eq}(\tau-u) 
\eea
and, after integarting by parts~:
\be
 - \frac{1}{T} \(( C_{eq}(t')C_{eq}^{p-1}(t) -
C_{eq}(\tau) \))
+\frac{1}{T} \int_\tau^{t} du \  C_{eq}^{p-1}(u)  C_{eq}'(u-\tau) 
- \frac{1}{T} \int_0^\tau du \  C_{eq}^{p-1}(\tau -u)  C_{eq}'(u).
\ee
We finally obtain
\bea
\frac{\partial C_{eq}}{\partial \tau} &=& - \((T +\frac{p}{2T} \))
C_{eq}(\tau) - \frac{p}{2T} \(( C_{eq}(t')C_{eq}^{p-1}(t) -
C_{eq}(\tau) \)) \\ 
&-& \frac{1}{T} \int_0^\tau du \  C_{eq}^{p-1}(\tau -u)  C_{eq}'(u)
+ \frac{p}{2T} C_{eq}^{p-1}(t)C_{eq}(t'),
\eea
which is equation (\ref{eqas}).
 
Remark~: it is possible to treat in the same way the equation for
the response function~; this yields the derivative
of (\ref{eqas}) with respect to  $\tau$.
 
\subsubsection{T dif\/ferent from T'}
 
For T dif\/ferent from T', there is a transient regime.
We therefore take $t$ and $t'$ large, i.e. the one-time quantities
have reached their limiting values~:
\bea
C(t,0) &=& C(t',0) = l \\
\mu(t) &=& \mu_{\infty}=T +  \frac{p}{2T} (1-C_{\infty}^p)
+\frac{p}{2T'} l^{p} .
\eea
 
The equation (\ref{as2}) is obtained in the same way as for
$T=T'$. The limit $\tau \to \infty$ yields
\be
0 = - \(( \mu_{\infty} - \frac{p}{2T} \)) C_{\infty}
- \frac{p}{2T} C_{\infty}^{p-1} (C_{\infty}-1) 
- \frac{p}{2T}C_{\infty}^p + \frac{p}{2T'} l^p,
\ee
or
{\boldmath
\be
\colorbox{grey0.95}{\makebox{$\displaystyle
T C_{\infty} = \frac{p}{2T'} l^p (1-C_{\infty}) + \frac{p}{2T}
C_{\infty}^{p-1} (1-C_{\infty})^2  .
\label{eqcinfini}
$}}
\ee}
Taking the limit of $t$ going to inf\/inity in the equation
for $C(t,t')$ with $t'=0$ (\ref{eqTT}), we obtain~:
\bea
0 &=&- \mu_{\infty} l + \frac{p(p-1)}{2}
\lim_{t \to \infty} \int_0^t C_{eq}^{p-2}(t-s)r_{eq}(t-s)C(s,0)
+ \frac{p}{2T'} l^{p-1} \\ \nonumber
&=& 
- \mu_{\infty} l - \frac{p\ l}{2T}\((C_{\infty}^{p-1} -1 \))
+ \frac{p}{2T'} l^{p-1}
\eea
and f\/inally~:
{\boldmath
\be
\colorbox{grey0.95}{\makebox{$\displaystyle
l^{p-2} = \frac{2T T'}{p (1-C_{\infty})} .
\label{eql}
$}}
\ee}
 
Let us consider the TAP states that dominate the partition function at
$T'$~; we note
\be
q=q(E^{0}_{T'},T) \mbox{\ \  et \ \ }q_1=q(E^{0}_{T'},T') .
\ee
Equations (\ref{qtapeq}) and ({\ref{qtap}) show that $q$  and $q_1$
verify~:
\be
\frac{p}{2T'^2}q_1^{p-2} (1-q_1) =1, \  
\frac{1-q}{T}q^{\frac{p}{2}-1} = \frac{1-q_1}{T'}q_1^{\frac{p}{2}-1},
\ee
so that
\be
q_1 = 1 - \frac{p}{2T^2} (1-q)^2 q^{p-2}
\label{q1fctionq}
\ee
If we now take
$X= 1 - \frac{p}{2T^2} (1-C_{\infty})^2 C_{\infty}^{p-2}$,
the equation for $C_\infty$ (coming from
(\ref{eqcinfini}) and (\ref{eql})) can be written~:
\be
X - \(( \frac{p}{2} (1-X) \))^{-1/(p-2)} T'^{2/(p-2)} = 0,
\ee
therefore $X=q_1$ and~:
{\boldmath
\be
\colorbox{grey0.95}{\makebox{$\displaystyle
C_\infty = q(E^{0}_{T'},T).
$}}
\ee}
 
The energy is (\ref{energie})~:
\be
E_\infty = \frac{T- \mu_{\infty}}{p} =
\frac{1}{2T}( C_{\infty}^p - 1 ) - \frac{l^p}{2T'}
\ee
(remark that for $T=T'$, we get back, with $l=C_{\infty}$,
$E_\infty = -\frac{1}{2T}$),
i.e.~:
\be
E_\infty = -\frac{T C_{\infty}}{p (1-C_{\infty})} 
+ \frac{C_{\infty}^{p-1}}{2T} -\frac{1}{2T}.
\label{einfini}
\ee
 
the energy at $T$ of the TAP states dominating the statics at $T'$
is~:
\be
E(E^{0}_{T'},T) = q^{\frac{p}{2}} E^{0}_{T'} - \frac{1}{2T} \((
(p-1) q^p - p  q^{p-1} +1 \))
\ee
which is identical to (\ref{einfini}) (this identity is obtained
using (\ref{q1fctionq}) and (\ref{en0tap}))~:
{\boldmath
\be
\colorbox{grey0.95}{\makebox{$\displaystyle
E_\infty = E(E^{0}_{T'},T).
$}}
\ee}

\end{document}